\documentclass[%
 reprint,
 amsmath,amssymb,
 aps,
]{revtex4-2}
\usepackage{verbatim}
\usepackage{verbatim,upgreek}
\usepackage{chngcntr}
\usepackage{graphicx}
\usepackage{dcolumn}
\usepackage{bm}

\begin{document}

\preprint{APS/123-QED}

\title{Electric-Field Programmable Spin Arrays for Scalable Quantum Repeaters}

\author{Hanfeng Wang$^{1,2}$}

\author{Matthew E. Trusheim$^{1,3}$}
 \email{mtrush@mit.edu}

\author{Laura Kim$^{1,2}$}
 
\author{Dirk R. Englund$^{1,2}$}
 \homepage{englund@mit.edu}
 
\affiliation{%
$^{1}$ Research Laboratory of Electronics, M.I.T, 50 Vassar Street, Cambridge, MA 02139, USA\\
$^{2}$ Department of Electrical Engineering and Computer Science, M.I.T., Cambridge, MA 02139, USA\\
$^{3}$ U.S. Army Research Laboratory, Sensors and Electron Devices Directorate, Adelphi, MD 20783, USA}%

\begin{abstract}

Large scale control over thousands of quantum emitters desired by quantum network technology is limited by power consumption and cross-talk inherent in current microwave techniques. Here we propose a quantum repeater architecture based on densely-packed diamond color centers (CCs) in a programmable electrode array. This `electric-field programmable spin array' (eFPSA) enables high-speed spin control of individual CCs with low cross-talk and power dissipation. Integrated in a slow-light waveguide for efficient optical coupling, the eFPSA serves as a quantum interface  for optically-mediated entanglement. We evaluate the performance of the eFPSA architecture in comparison to a routing tree design and show an increased entanglement generation rate scaling into the thousands of qubits regime. Our results enable high fidelity control of dense quantum emitter arrays for scalable networking.

\end{abstract}

\maketitle

\section{\label{sec:level1}Introduction\protect}

Future quantum repeaters or modular quantum computers will need to manage large numbers of multiplexed memory qubits with efficient local operations. Solid-state artificial atoms such as color centers (CCs) in diamond are promising quantum memories \cite{ruf2021quantum,bhaskar2017quantum}. Precision control of the electronic spin ground state of CCs presently relies on AC magnetic fields \cite{fuchs2011quantum,zhang2017selective,hong2013nanoscale,koehl2011room}. Developing architectures for spatially multiplexed microwave control with sufficiently low power dissipation and cross-talk remains an open challenge. Here, we propose a fundamentally different approach that uses the spin-electric field coupling of the CC ground state. We show that electric-field-based spin control offers lower power dissipation and cross-talk, as well as compatibility with integrated circuit (IC) platforms likely needed for scaling. With an efficient optical interface and all-to-all connectivity, our platform can be integrated to allow scalable entanglement generation. 
 
We consider a programmable array of electrodes positioned around arrays of CCs in a diamond waveguide. This `electric field programmable spin array' (eFPSA) architecture has three key elements: 1. A quantum memory based on the diamond nitrogen-vacancy (NV) center, which has already been used for optical entanglement distribution across as many as three qubits \cite{pompili2021realization, humphreys2018deterministic, bernien2013heralded}. 2. An efficient optical interface through a slow-light photonic crystal (PhC) waveguide  enabling $\sim 25\times $  Purcell enhancement of the NV's coherent transition. 3. An electrode array positioned along individual qubits in the waveguide. We estimate that the eFPSA enables $\sim100$ ns-duration spin rotations, as well as $\sim600$ GHz-range DC Stark shift tuning of CC optical transitions. 

The article is organized as follows. Sect.~\ref{sec1} introduces the eFPSA and estimates achievable gate performance, focusing chiefly on duration and cross-talk. Sect.~\ref{sec2} describes the co-designed PhC waveguide to achieve high emitter-waveguide coupling with low optical loss, as well as dynamical optical tunability by Stark shift.  Sect.~\ref{sec3} combines the elements of two previous sections to show a quantum repeater architecture enabled by the eFPSA, illustrating how the eFPSA can mediate local qubit interactions and multiplexed quantum network connectivity. 

\section{\label{sec:level1}Results\protect}

\subsection{\label{sec:level2}Localized single-qubit control } \label{sec1}

Fig.~\ref{fig:1}(a) shows the eFPSA design. It consists of a single-mode diamond waveguide with a centered NV array, placed onto high-index dielectric fins between an electrode array. We use HfO$_2$ for the dielectric fins as it has a high dielectric constant of 23 in the radio frequency range and a relatively low index of 1.9 in the optical range \cite{lin2002dielectric,al2004optical}. This allows it to concentrate the low frequency electric field required for spin coupling while guiding the optical mode. We assume indium tin oxide (ITO) as the electrode material to minimize optical loss. Finally we use SiO$_2$ as a low-index substrate.   

\begin{figure*} \label{fig:1}\includegraphics[width = 1.0\textwidth]{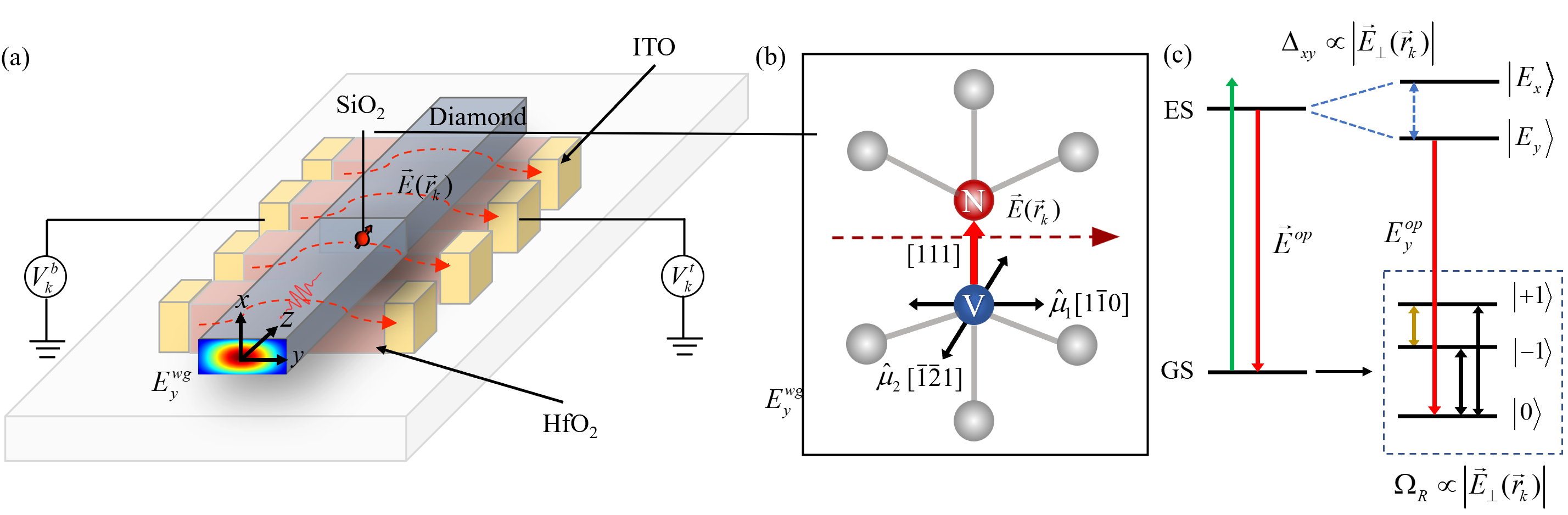}
\caption{\label{fig:1} (a) An exemplary eFPSA implementation consists of a diamond waveguide containing NV centers positioned on an array of high-index dielectric fins between ITO electrodes connected by HfO$_{2}$. Electric fields $\vec{E}(\vec{r}_k)$ are generated by top (bottom) contacts at potentials $V_{k}^t$($V_{k}^b$) for NV spin control. The NV fluorescence couples to the waveguide mode $\vec{E}_{z}^{wg}$. Here the $\hat{x}=[001]$, $\hat{y}=[1\bar{1}0]$, $\hat{z}=[110]$. (b) The NV center in diamond in the [111] direction with transition dipoles $\vec{\mu}_1=[1\bar{1}0]$ and $\vec{\mu}_2=[\bar{1}\bar{2}1]$. (c) The Zeeman shift due to the magnetic field splits the $m_s=\pm1$ sublevels, while an electric field splits the excited state levels $|E_x\rangle$ and $|E_y\rangle$ by $\Delta_{xy}$ via the DC Stark effect (left). The black (yellow) line indicates the Rabi transition between $|\pm1\rangle\leftrightarrow|0\rangle$ ($|+1\rangle\leftrightarrow|-1\rangle$) driven by a resonant transverse electric field. }
\end{figure*}

As illustrated in Fig.~\ref{fig:1}(a), we consider an array of NV spin memories approximately at a periodic spacing $a$, i.e., the $k$th NV has a position $\vec{r}_k=ka\hat{z}+\vec{\delta}_k$, where $|\vec{\delta}_k|/a\ll 1$. The relevant interaction between an NV ground state spin and an electromagnetic  field $\vec{E}$ and $\vec{B}$ is captured in the Hamiltonian: \cite{dolde2011electric, doherty2012theory, udvarhelyi2018spin} 
\begin{equation} \label{eq:1}
\begin{aligned}
    H_{\vec{E},\vec{B}}&/h = \gamma \vec{S}\cdot \vec{B}+ d_{\perp}[\{S_{x'},S_{z'}\}E_{x'}+\{S_{y'},S_{z'}\}E_{y'}]\\
    &+d_{\parallel}S_{z'}^2E_{z'}+d_{\perp}'[(S_{y'}^2-S_{x'}^2)E_{x'}+\{S_{x'},S_{y'}\}E_{y'}]
\end{aligned}
\end{equation}
where $d_{\perp}=|\hat{z}'\times(\hat{z}'\times \vec{d})|= $ 0.35 Hz cm/V ($d_\parallel=\hat{z}'\cdot \vec{d}$ = 17 Hz cm/V) denotes the perpendicular (parallel) part of spin-electric field susceptibility \cite{dolde2011electric},  $h$ the Planck constant, $\mathbf{S}$ the electron spin operator, and $\gamma$ the gyromagnetic ratio. $d_{\perp}'$ has not been quantified experimentally or theoretically, but is  estimated near $1/50~d_{\perp}$ \cite{doherty2011negatively, udvarhelyi2018spin}. Here we use the primed coordinates $(x',y',z')$ to indicate the coordinates relative to the NV aligned along the $z'$ axis.

\textbf{Spin-electric coupling -} We now consider an external electric driving field $\vec{E}_{\pm 1\leftrightarrow 0}$ ($\vec{E}_{+1\leftrightarrow -1}$) resonant with the  $|\pm1\rangle\leftrightarrow|0\rangle$ ($|+1\rangle\leftrightarrow|-1\rangle$) transitions, which are non-degenerate under a small bias magnetic field along the $z'$-axis, as shown in Fig.~\ref{fig:1}(c). From the Schrödinger equation, the Rabi frequency of coherent driving on the NV ground state triplet is:
\begin{equation} \label{eq:2}
\begin{aligned}
    h\Omega^{\pm1\leftrightarrow 0}_{R}&=\frac{1}{\sqrt{2}}d_{\perp}'|\vec{E}_{\perp}(\vec{r}_k)|\\
    h\Omega^{+1\leftrightarrow -1}_{R}&=d_{\perp}|\vec{E}_{\perp}(\vec{r}_k)|
\end{aligned}
\end{equation}
where $\vec{r}_k$ indicates the positions of NVs shown in Fig.~\ref{fig:1}(a) and $\vec{E}_\perp(\vec{r}_k)$ is the component of electric field perpendicular to NV axis. As shown in Fig.~\ref{fig:1}(b), we choose $\vec{\mu}_1=[1\bar{1}0]$ and $\vec{\mu}_2=[\bar{1}\bar{2}1]$ as basis vectors for the plane perpendicular to NV axis, i.e. $\vec{E}_\perp(\vec{r}_k)=(E_{\vec{\mu}_1}(\vec{r}_k),E_{\vec{\mu}_2}(\vec{r}_k)$) where $\vec{\mu}_1$ and $\vec{\mu}_2$ are also the axes of NV optical transitions $E^{op}_y$ and $E^{op}_x$, respectively \cite{epstein2005anisotropic}.   
\begin{figure}[b]

\includegraphics[width = 0.48\textwidth]{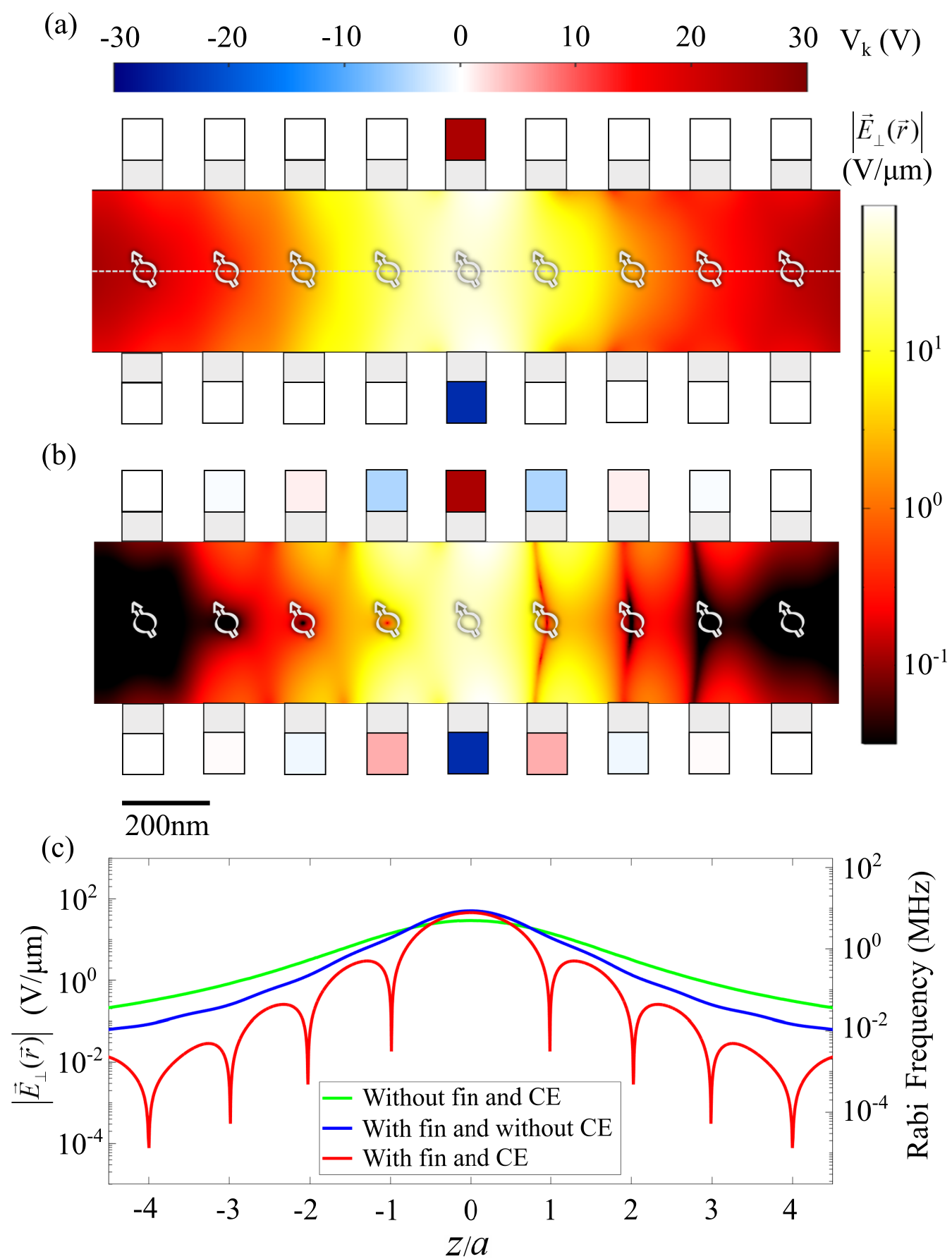}\caption{\label{fig:2} (a). Electric field component $|\vec{E}_\perp|$ at the waveguide midplane $x=0$ with a 50 V (2 GHz)-signal applied to the central electrode, while all other electrodes are connected to the ground. (b). Electric field component $|\vec{E}_\perp(x=0,y,z)|$ with E-field concentrator HfO$_2$ fin and cross-talk elimination (CE). (c). Electric field component $|\vec{E}_\perp(x=y=0,z)|$ along the white dotted line in (a), and corresponding Rabi frequency of the $|+1\rangle\leftrightarrow |-1\rangle$ transition. }
\end{figure}

\textbf{Electric field-driven gate speed - } Absent other experimental noise, the single qubit gate fidelity is limited by the inhomogeneous dephasing time $T_2^*\sim 10$ $\upmu s$ \cite{bauch2018ultralong}.  For a pure superposition state, the fidelity of a $\pi$-rotation at Rabi frequency $\Omega_R$ under this dephasing process is $F_{\mathrm{dephasing}} = 1/2(1+\exp(-1/2\Omega_R T_2^*))$. Considering random pure states uniformly-distributed on the Bloch sphere, the average fidelity reaches above 0.99 with a Rabi frequency of $1.7$ MHz. A resonant electric field of $10^7$ V/m is needed for reaching this gate fidelity. In our geometry, this requirement is met for a $\sim 10$ V potential difference, which is compatible with modern integrated circuits technology such as complementary metal-oxide semiconductor (CMOS) platforms. We estimate that electric field-driven Rabi frequencies can reach $\Omega^{+1\leftrightarrow -1}_{R}\sim0.13$ GHz and $\Omega^{\pm1\leftrightarrow 0}_{R}\sim$ 1.9 MHz, limited by diamond's dielectric strength $E_{\mathrm{bd}}^{\mathrm{dmd}}\sim 2\times 10^3$ V/$\upmu$m \cite{landstrass1993device,volpe2010extreme} and HfO$_{2}$ dielectric strength $E_{\mathrm{bd}}^{\mathrm{HfO}_2} \sim 1.6\times10^3$ V/$\upmu$m \cite{kuo1992study} at a separation of hundreds of nm. 
\begin{table}[b]
\caption{\label{tab:table1}%
eFPSA Parameters
}
\begin{ruledtabular}
\begin{tabular}{lcdr}
 & Symbol &  \mathrm{Value} \\
\colrule
Electrode spacing & $a$ & 0.183~ \upmu \mathrm{m} \\
Diamond waveguide thickness & $h_{wg}$ & 0.364~ \upmu \mathrm{m} \\
Diamond waveguide width & $w_{wg}$ & 0.091~ \upmu \mathrm{m} \\
HfO$_2$ length & $l_{\mathrm{HfO_2}}$ & 0.500~ \upmu \mathrm{m} \\
HfO$_2$ width & $w_{\mathrm{HfO_2}}$ & 0.091~ \upmu \mathrm{m} \\
HfO$_2$ thickness & $h_{\mathrm{HfO_2}}$ & 0.273~ \upmu \mathrm{m} \\

\end{tabular}
\end{ruledtabular}
\end{table}

\textbf{Electric field-driven gate fidelity - } Fig.~\ref{fig:2}(a) plots the $|\vec{E}_{\perp}(x,y,z=0)|$ electric field component when a voltage $V_k=V_k^t-V_k^b=50$ V is applied in eFPSA with parameters in Table I, obtained from Maxwell's equations using COMSOL Multiphysics. Here $V_k^{t(b)}$ indicates the voltage on $k$th top (bottom) electrode. As the NV is in the [111] axis and each field component $E_x$, $E_y$ and $E_z$ are symmetric along the $z$-axis. However, $E_z$ is odd whereas $E_x$ and $E_y$ are even functions. Therefore for a NV along [111] axis, $E_\perp=\sqrt{E_y^2+(\cos54.7^{\circ} E_x+\sin54.7^{\circ} E_z)^2}$ does not have symmetry along $z$-axis. However, when $y=0$ (shown in the white dotted line of Fig. 2(a)), $E_z(z)=0$ and $E_\perp$ is symmetric along $z$. 

The closest separation between individually controllable NVs in an array is limited by the cross-talk between the target NV at location $\vec{r}_k$ and a neighboring NV at $\vec{r}_{k+1}$. During a $\pi$-pulse on qubit $\vec{r}_{k}$, there is an undesired rotation on $\vec{r}_{k+1}$. We evaluate the cross-talk fidelity $F_{\mathrm{C}}$ by comparing $\mathbf{R}(\vec{r}_{k+1})$ and desired identity operation $\mathbf{I}(\vec{r}_{k+1})$. This fidelity can be expressed as
\begin{equation}
F_{\mathrm{C}}(\mathbf{R}(\vec{r}_{k+1}),\mathbf{I}(\vec{r}_{k+1}))=\left(\mathrm{tr}\sqrt{\sqrt{\rho}\sigma\sqrt{\rho}}\right)^2
\end{equation}
where 
\begin{equation}
\begin{aligned}
\rho&=\mathbf{R}(\vec{r}_{k+1})|\psi_0(\vec{r}_{k+1})\rangle\langle\psi_0(\vec{r}_{k+1})|\mathbf{R}^\dagger(\vec{r}_{k+1})\\
\sigma&=\mathbf{I}(\vec{r}_{k+1})|\psi_0(\vec{r}_{k+1})\rangle\langle\psi_0(\vec{r}_{k+1})|\mathbf{I}^\dagger(\vec{r}_{k+1})
\end{aligned}
\end{equation}
and $|\psi_0(\vec{r}_{k+1})\rangle$ is the initial quantum state of NV at location $\vec{r}_{k+1}$. For the profile shown in Fig.~\ref{fig:2}(a), $F_{\mathrm{C}
}^{\mathrm{fin}}=0.92$. The field confinement provided by the HfO$_{2}$ fins results in a significant improvement over bare electrodes, where $F_C^\mathrm{bare}=0.69$ (see Appendix).

\textbf{Cross-talk elimination - }To further reduce the cross-talk, we use our individual control over $2N$ voltages $V=\{V_k^{t(b)}\}$ to eliminate the electric field at the locations of the non-target qubits. Electric field applied on each qubit $\vec{E}_\perp=\{E_{\vec{\mu}_1}(\vec{r}_k),E_{\vec{\mu}_2}(\vec{r}_k)\}$ has a linear dependence with the voltage by $\vec{E}_\perp=GV$, where $G_{ij}$ is the linear map between $E_i$ and $V_j$, computed by COMSOL Multiphysics. $V$ are then chosen to minimize the cross-talk by $V_{\mathrm{CE}}=G^{-1}\vec{E}_{\mathrm{tar}}$. For the case of a single qubit gate on a target NV at location $i$, we set $\vec{E}_{\mathrm{tar}}=\vec{E}_{\perp,\mathrm{tar}}^T\otimes_{\mathrm{K}}\delta_{ik}$ and $\vec{E}_{\perp,\mathrm{tar}}$ is the AC electric field applied on the target NV. Since the number of independent degrees of freedom is equal to the number of electric field values to be minimized, this inversion is possible. As shown in Fig.~\ref{fig:2}(b,c), the cross-talk elimination process creates low electric field on non-target NV positions, increasing the cross-talk fidelity to $F_{\mathrm{CE}}>0.999$. Now, the total infidelity is mainly caused by dephasing rather than cross-talk. This procedure can be used for any arbitrary operations over all NVs by the above procedure, choosing a specific $\vec{E}_{\perp,\mathrm{tar}}$.

\textbf{Heat-load for electric field- vs magnetic field-based spin control}. Heat dissipation is critically important in cryogenic environments, where cooling power is limited and heating can degrade performance. We approximate the low-temperature stage power consumption of electric field-based coherent control by modeling the eFPSA as a capacitance $C$ with a parallel resistance $R$ in series with a wire (resistance $R_w\sim 10^{-2}~\Omega$) inside the cryogenic environment (circuit details in Appendix).  We choose the figure of merit as the energy per spin $\pi$-pulse that is deposited at the low-temperature stage. In our design, the eFPSA acts like an open circuit and almost all the power is reflected back to the high temperature region. For $|+1\rangle\leftrightarrow|-1\rangle$ transition, the energy dissipation in the cryostat is given by: 
\begin{equation}
    J_{E} = \frac{1+\omega^2C^2R_wR}{R}\frac{\Lambda^2\Omega_R}{2d_\perp^2}
\end{equation}
where $\omega$ is the frequency of the AC electric field and $\Lambda\sim 1~\upmu$m is the characteristic length that relates the applied voltage on the eFPSA and electric field at the NV positions. In our geometry, $C=2.8\times10^{-17}$ F is the circuit capacitance simulated by COMSOL and $R\sim 10^{20}~\Omega$ is the resistance of the HfO$_2$ calculated from thin film resistivity \cite{yakovkina2005preparation}. The energy dissipation in the cryostat per $\pi$-pulse $J_E$ for a Rabi frequency $\Omega_R=2$ MHz is $1.1\times10^{-21}$ J. A second figure of merit is the dissipation ratio between electric field and magnetic field-based driving with the same Rabi frequency,  $J_E/J_B$. Here we take the magnetic circuit to be the bare wire with a resistance $R_w$ and a capacitance $C_w$, with the NV positioned at a distance $\Lambda$ from the wire. Then the ratio is given by:
\begin{equation} \label{eq:5}
    \frac{J_E}{J_B}\sim\frac{\mu_0^2\gamma^2}{4\pi^2d_\perp^2}\frac{1+\omega^2C^2R_wR}{R_wR}
\end{equation}
where $\mu_0$ is the vacuum permeability. Here $J_E/J_B = 3.4\times 10^{-6}$, suggesting the power dissipation for electric field control is 6 orders of magnitude less than that for magnetic field control. For driving single quantum transition $|0\rangle\leftrightarrow|\pm1\rangle$, $J_E/J_B = 1.7\times 10^{-2}$. In the real case, we need to consider the leakage current. For a $\sim$ P$\Omega$ leakage resistance \cite{matsuda1994measurements}, we will have $J_E/J_B = 2.7\times 10^{-5}~ (1.4\times10^{-1})$ for $|\pm 1\rangle\leftrightarrow|0\rangle$ ($|+1\rangle\leftrightarrow|-1\rangle$) transitions.

\subsection{\label{sec:level2}Efficient coupling of an NV array to a slow-light waveguide} \label{sec2}

The entanglement rate of NV centers relies on the spin-photon coupling efficiency, which is given by \cite{arcari2014near,rao2007single}:
\begin{equation} \label{eq:6}
    \beta = \frac{F_P\cdot\Gamma_{\mathrm{wg0}}}{F_P\cdot\Gamma_{\mathrm{wg0}}+\Gamma_{\mathrm{others}}}
\end{equation}
where $\Gamma_{\mathrm{wg0}}$ is the decay rate of spin-entangled transition in the absence of any optical structures, and $\Gamma_{\mathrm{others}}$ the total rate of all other decay mechanisms. Here we focus on the $E_y^{op}$ transition of the NV center with a frequency $\nu_{0}$.

The spontaneous emission rate of a quantum system can be enhanced by modifications in its electromagnetic environment. The rate of transition of an emitter scales with the final density of states to first-order approximation according to Fermi’s golden rule. Slow-light waveguide structures produce a photonic bandgap, resulting in a small group velocity near the band edge. As a result, an emitter placed in the mode maximum of a slow-light waveguide experiences a large enhancement in the local density of electromagnetic states, and its rate of transition is enhanced by the Purcell factor \cite{javadi2018numerical,rao2007single}.

\begin{figure}[b]
\includegraphics[width = 0.48\textwidth]{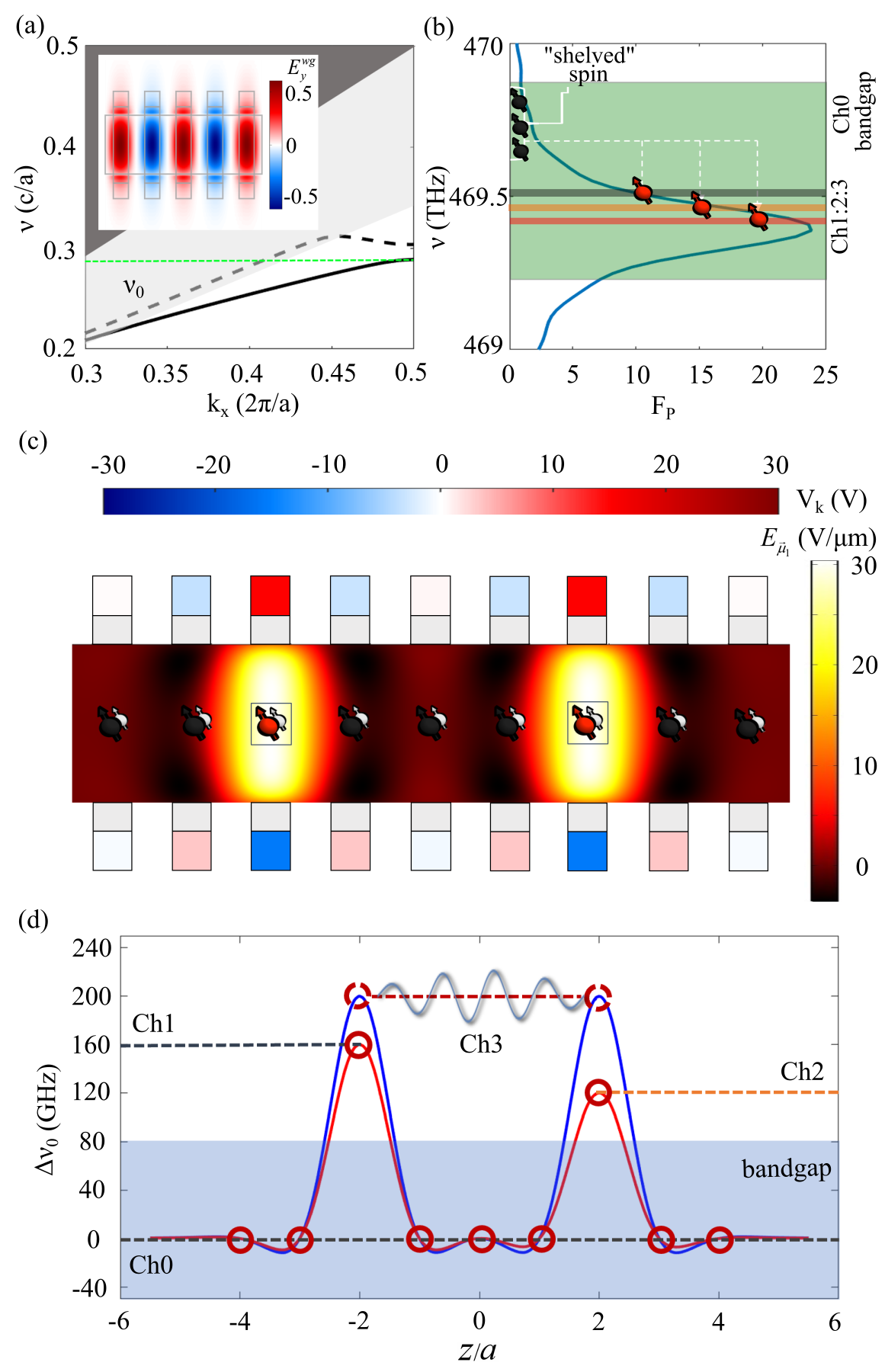}
\caption{\label{fig:3} (a). Photonic TE band structure of the eFPSA slow-light waveguide using the parameters shown in Table 1. The dark (light) shaded regions indicate the light cone for propagation in free space (substrate). The inset shows the $y$-component of the electric field at the midplane of the diamond. (b). Purcell factor for the eFPSA with a finite number of periods $N=100$ near the bandgap. Green shading indicates the Stark tuning range, and the horizontal shading indicates non-interacting frequency channels. (c). Electric field profile for placing two NVs at $j=\pm 2$ on resonance. (d). NV Stark shift versus positions for two voltage configurations. The blue curve shows the Stark shift for the electric field profile shown in (c). The red curve shows another voltage setting where two emitters are set in non-interacting channels. }
\end{figure}

Conveniently, the RF-field-concentrating fin structures also provide a periodic dielectric perturbation, forming a slow-light mode in the optical band. Fig.~\ref{fig:3}(a) indicates the TE-like modes of the slow-light waveguide with the parameters shown in Table 1. By coupling the NV transition at $\nu_{0}$ to the slow-light region, we can thus funnel the coherent emission into waveguide modes near wave-vector $k_x(\nu_{0})$, as shown in Fig. 3(a). From finite difference time domain (FDTD) simulations (Lumerical), we obtain a maximum Purcell factor of $F_{P\mathrm{max}}=25$ for an NV in the [111] direction placed on the mid plane of the diamond waveguide when the number of periods is 100 (Fig.~\ref{fig:3}(b)).

The total photon collection efficiency out of the waveguide is given by $\eta_{wg}=\beta \exp(-t_{wg}N)$, where $t_{wg}$ is the waveguide transmission from the emitter to the waveguide facet. Assuming a NV Debye-Waller factor of $DW=0.03$ \cite{doherty2011negatively} and  we use the relation $\Gamma_{\mathrm{others}}/\Gamma_{\mathrm{total}}=1-DW$ to calculate $\beta = 25\%$ following Eq. 8. \cite{wolters2010enhancement,santori2010nanophotonics} The transmission efficiency $t_{wg}\sim 4\times 10^{-3}$ dB/period estimated from FDTD simulations.

\textbf{Spectral addressing by Stark shift - } To selectively couple the waveguide propagating modes to a specific NV center in the array, we use the electrodes for another function: to induce a local Stark shift via an applied DC electric field. Under an applied electric field, the emitter's natural $E_y^{op}$ transition at $\nu_{0}$ is shifted to $\nu_{0}+\Delta\nu_{0}$, where $\Delta\nu_{0}$ is given by: \cite{bassett2011electrical}
\begin{equation} \label{eq:3}
h\Delta\nu_0=\Delta\mu_{\parallel}E_{\parallel}-\frac{\sqrt{2}}{2}\mu_{\perp}\sqrt{E_{\vec{\mu}_1}^2+E_{\vec{\mu}_2}^2}
\end{equation}
where $E_{\parallel}$ is the electric field along [111], $\Delta\mu_{\parallel}=\mu_{\parallel}^{\mathrm{GS}}-\mu_{\parallel}^{\mathrm{ES}}\sim 1.5$ D is the parallel dipole moment difference between excited states and ground states. $\mu_{\perp}\sim$ 2.1 D is the perpendicular component of electric dipole moment. Here we choose $E^{op}_y$ transition to avoid the depopulation and mixing of excited states at large applied fields \cite{fronik2018homogeneous}.

The maximum tuning range using this effect is $\sim$ 600 GHz assuming an applied electric field of $10^8$ V/m in the eFPSA architecture, indicated by the green shaded area in Fig.~\ref{fig:3}(b). This corresponds to tuning across the full range of the slow-light Purcell enhancement and into the waveguide bandgap. As the Purcell-enhanced NV ZPL transition linewidth is $\sim$ 100 MHz, the wide range of the Stark tuning allows multiple frequency channels where NVs are selectively placed into distinct bands that can be individually addressed in the frequency domain. Three channels (Ch1$-$3) spaced by 40 GHz and an off-resonant channel (Ch0) are indicated in Fig.~\ref{fig:3}(b). The Purcell enhancement across Ch1$-$3 is maintained at $\sim$ 10, while the large spacing suppresses interactions (photon absorption) between channels \cite{bersin2019individual}.

The ability for eFPSA to reconfigure the electric field profile locally allows for arbitrary and independent configuration of NV optical transitions. Unlike in Sec. I where the $E_{\parallel}$ is neglected, here we need 3$N$ degree of freedoms to control components $E_{\parallel}$, $E_{\vec{\mu}_1}$ and $E_{\vec{\mu}_2}$ for each of the $N$ NVs. However, we can use symmetry $V_{t}+V_{b}=0$ to set $E_\parallel= E_{\vec{\mu}_2}= 0$ in the ideal case. The remaining $N$ degrees of freedom can be used to set $E_{\vec{\mu}_1}$ for all the NVs. For example, we compare two configurations in Fig. 3(d). The blue curve shows the frequency shift for an electric field profile in Fig.~\ref{fig:3}(c), resulting in two NVs on resonance. The red curve shows the frequency shift for a different voltage setting, where two NVs are in different channels without interaction. In both cases, all other NVs are off-resonant in Ch0. The ability to dynamically control the NV transition frequency via electrical control can then be used to perform individual emitter initialization and readout, and to reconfigure quantum network connectivity as described below.

\subsection{\label{sec:level2}Quantum repeater performance} \label{sec3}

\begin{figure*}
\includegraphics[width = 0.92\textwidth]{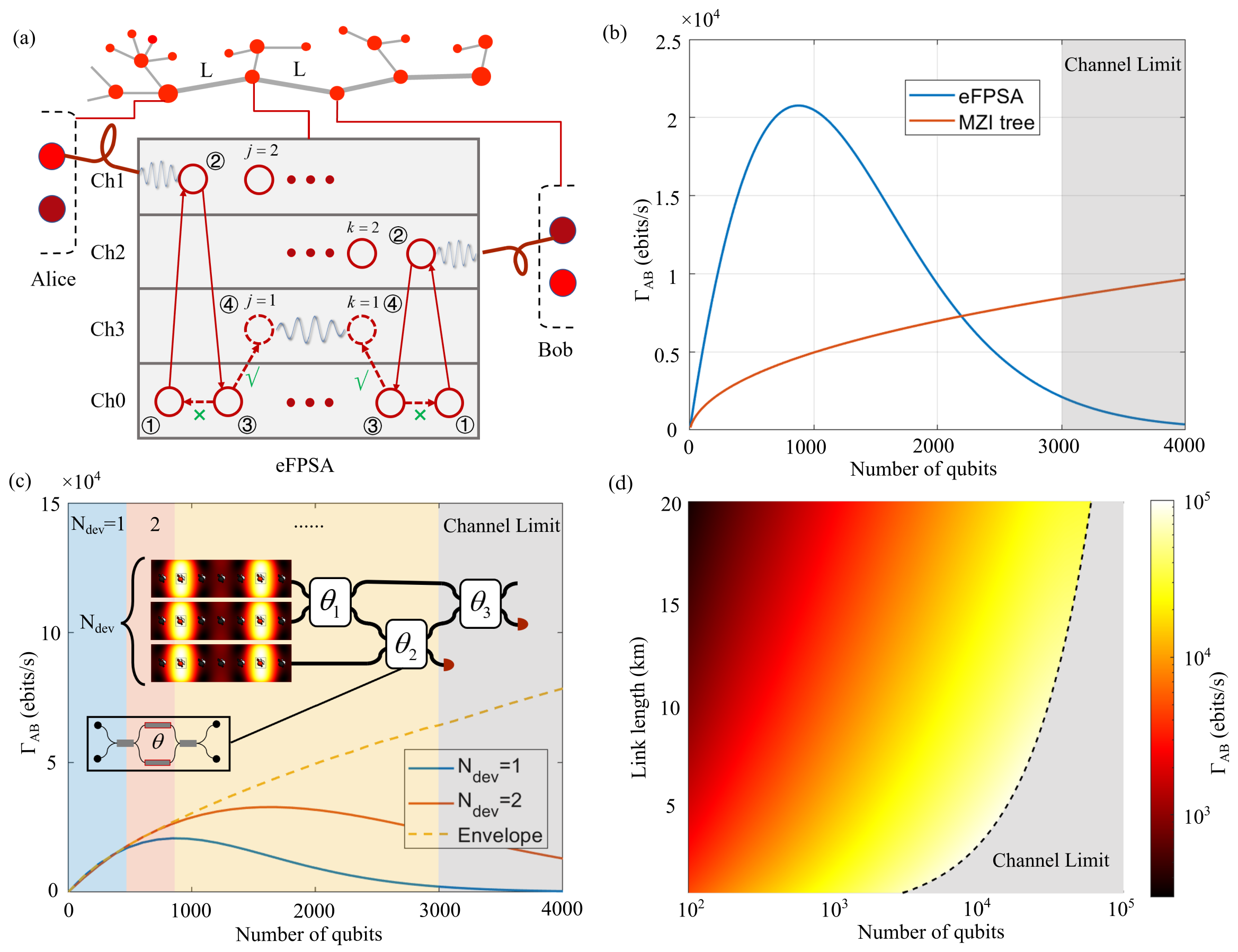}
\caption{\label{fig:4} (a). Quantum repeater architecture containing quantum nodes (red) connected by channels (gray). Here we consider a 3-node link containing Alice, Bob and a central eFPSA. NVs are Stark-shifted to frequency channels (Ch0-Ch3) at different points in the protocol indicated by step 1-4. (b). $\Gamma_{AB}$ over a L $=1$ km channel as a function of the number of qubits for eFPSA (red) and MZI tree (blue) architectures. Above 3000 qubits, the entanglement procedure is limited by the channel capacity. (c). $\Gamma_{AB}$ as a function of the number of qubits for a hybridization of MZI and eFPSA architecture. (d). Optimal $\Gamma_{AB}$ for different link length and number of qubits. The dotted line shows the largest $\Gamma_{AB}$ for different link length.}
\end{figure*}

We next consider the application of the eFPSA as a quantum repeater to generate Bell pairs $|\psi_{AB}\rangle$ between two memory qubits at Alice and Bob (A and B), as illustrated in Fig.~\ref{fig:4}(a). The repeater protocol has two steps: 

(1) Distant entanglement between A(B) and electron spin $|j_e\rangle$($|k_e\rangle$) in the eFPSA via a heralded single-photon scheme, which has previously been demonstrated for NV centers
\cite{humphreys2018deterministic}, followed by swapping to the $^{15}N$ nuclear spin $|j_n\rangle$($|k_n\rangle$). For each link, we assume a length-$L$ noiseless channel with transmission $\eta=\exp(-\gamma L)$, where $\gamma=0.041$ km$^{-1}$ \cite{lee2020quantum}. Each entanglement attempt has a success probability of $p_1=2\alpha\eta p_d p_c \eta_{wg}$, where $p_d = 0.83$ ($p_c=0.33$) is the detection (quantum frequency conversion, if necessary) efficiency \cite{yu2020entanglement,le2016high}. We conservatively assume a lower $F_P = 10$ ($\beta = 25\%$) to avoid high loss and fabrication sensitivity in the regime of high group index \cite{o2010loss}. Here we set $\alpha=0.01$ to keep the two-photon excitation error of this scheme below 1\% \cite{humphreys2018deterministic}. 

(2) Local entanglement swapping to generate entanglement between A and B. First, (2.i) CC electron spins $|j_e\rangle$ and $|k_e\rangle$ are Stark-shifted to the same optical transition and entangled via the two-photon Barrett-Kok scheme \cite{barrett2005efficient,bernien2013heralded} with success probability $p_2=(p_d \eta_{wg})^2/2$. Then, (2.ii) Bell measurements in the electron-nuclear spin basis of the memories $j$ and $k$ swap the local entanglement to distant entanglement of A and B after subsequent feed-forward. This step makes use of the eFPSA's all-to-all connectivity to realize a `quantum router' architecture \cite{lee2020quantum} that minimizes the latency (waiting time and associated decoherence) and local buffer size.

To evaluate the performance, we consider the entanglement rate $\Gamma_{AB}$ as the figure of merit, defined as the number of generated Bell pairs $|\psi_{AB}\rangle$ per second. For two qubits, $\Gamma_{AB}$ is the inverse of total time used for single pair entanglement generation. Parallel operations of $N$ pairs can increase this rate by a factor of $N$. The eFPSA time-multiplexes spin-photon entanglement to A and B, sending spin-entangled photons from different emitters in short succession. It is implemented by Stark-shifting the optical transition of selected color centers $j$ and $k$ from the bandgap to Ch1 and Ch2 (shown in Fig.~\ref{fig:4}(a)), while all other `unselected' color centers remain in the photonic bandgap. After an entanglement generation attempt, the NVs $j,k$ are tuned back into the bandgap and the process is repeated with the subsequent pair of NVs $j,k = j+1,k+1$ as shown in Fig. 4(a). In this way, time-multiplexing channels allow $N_{\mathrm{ch}}=t_{\mathrm{link}}/t_{\mathrm{ph}}$ qubits operate in parallel, where $t_{\mathrm{link}}$ is the heralding time for an optical pulse traveling photon in the fiber link and $t_{\mathrm{ph}}$ is the CC photon lifetime. For a $1$ km link and $10$ ns lifetime, we have $N_{\mathrm{ch}}\sim 300$ time-multiplexing channels. Additional frequency channels can further raise the number of time-frequency bins for distant entanglement generation \cite{spectrally}. The Stark shift tuning range, bandwidth of the slow-light effect, qubit linewidth and frequency multiplexer bandwidth limit the number of multiplexing channels. In our device, the regime of the purcell factor $>$ 10 has a bandwidth of $\sim 200$ GHz, setting the total frequency range. Here we assume a 20 GHz bandwidth for the frequency multiplexer based on the potential dense wavelength division multiplexing \cite{seyringer2012design}. Therefore, the time-frequency channel capacity in eFPSA is $10~ t_{\mathrm{link}}/t_{\mathrm{ph}}$. In the regime that we have fewer qubits than the channel capacity, each qubit pair can effectively generate entanglement independently. Above this number, qubits will compete for channel usage. 

The same dynamically tunable operations allow us to immediately attempt local entanglement as soon as distant entanglement is heralded. After a heralding signal from both A and B, we shift both qubits $j,k$ to Ch3 and generate local entanglement as described above. Due to localized independent electric field-based control, we can parallelize local entanglement generation (step 2) while simultaneously attempting entanglement over the long distance (step 1) links using other qubits (e.g. $j,k = j+1,k+1$) rather than requiring sequential operations. The entanglement rate is then mainly limited by the first step.


While increasing qubit number improves $\Gamma_{AB}$ linearly in the ideal case, each additional qubit adds exponential loss to the device as larger device size leads to transmission $\eta_{wg}\propto \exp(-t_{wg} N)$. The entanglement rate is given by $\Gamma_{AB}\propto N\exp(-t_{wg} N)$, as shown in the blue curve in Fig.~\ref{fig:4}(b) for parameters given in Table 1. The eFPSA reaches a maximum rate of $\sim 2\times 10^4$ ebits/s when the number of qubits is $\sim 800$, after which loss decreases rate exponentially with increasing number of qubits. The red curve shows the rate by a Mach-Zehnder interferometer (MZI) tree architecture as a comparison \cite{harris2017quantum}. Here we assume an efficiency 92\% per MZI \cite{joo2018cost, thomson2010low}. In the limit of large qubit number, the rate scales linearly with $0.92^{\mathrm{log}_2N}\propto N$. With very large qubit numbers, the MZI architecture could outperform eFPSA because it is not suffering from exponential loss. However, the number of parallel qubits is limited by the time-frequency multiplexing channel capacity. For a 1 km link, the channel capacity is 10 $t_{\mathrm{link}}/t_{\mathrm{ph}}\sim 3000$ shown in Fig. 4(b). In this regime, The eFPSA outperforms the MZI architecture by a factor of 4.

As an extension, we consider a hybridization of eFPSA and MZI tree architectures. As shown in Fig.~\ref{fig:4}(c), we divide the qubits into $N_{dev}$ eFPSAs connected by an MZI tree. $\Gamma_{AB}$ as a function of the number of qubits with different $N_{dev}$ is shown in Fig. 4(c). Taking the optimal $N_{dev}$ for each qubit number, we plot the maximum rate envelope shown in the dashed line. Instead of exponential decay, the optimal envelope asymptotically follows a linear scaling. In this scheme, the rate is limited by time-frequency multiplexing channels capacity shown in the gray region in Fig.~\ref{fig:4}(c). For different link lengths, the channel limit changes, resulting in varied maximum rate as shown in Fig.~\ref{fig:4}(d). The rate can be straightforwardly increased with additional frequency-multiplexing channels. Alternatively, a fixed number of memories $N$ can be used more efficiently in schemes with a midway entangled photon pair source \cite{jones2013high,kevinchen}, increasing the entanglement rate from $\propto \eta$ to $\sqrt{\eta}$.

\section{\label{sec:level1}Discussion\protect}

We comment briefly on (i) fabrication and (ii) qubit choices of the presented eFPSA blueprint. 

(i) Since the diamond can be placed on top of a periodic dielectric perturbation (e.g., by pick-and-place of diamond waveguides \cite{barclay2008pick,wan2020large}), enabling the substrate to be designed separately, a number of material choices are available. A potential approach is to produce high-index dielectric fins is to use atomic layer deposition of HfO$_2$ followed by lithography and a lift-off process \cite{liu2005ald}. Moreover, the substrate can be fabricated in CMOS to proximally position the required electrical contacts through a back-end-of-line metallization step. Custom CMOS processes have already been successfully demonstrated for NV spin control via microwave magnetic field interactions \cite{kim2019cmos}, whereas the electric field control should be easier as less current is required for a given Rabi frequency. Other materials with high dielectric constant such as barium titanate (BTO, $\varepsilon\sim7000$) would serve as attractive alternatives to HfO$_2$ as they could also provide electro-optic modulation of traveling modes \cite{abel2019large}.  


(ii) We considered the diamond NV center because of the reported spin-electric field coupling Hamiltonian and high dielectric strength of diamond -- but the NV has several drawbacks. Its rather low coupling strength $d_\perp'$ for electric field driving of the $|\pm1\rangle\leftrightarrow|0\rangle$ spin transition limits the Rabi frequency $f\propto d_\perp'$ that can be achieved without driving up the power dissipation $P_E\propto 1/d_\perp'^2$ or risking electrical breakdown. One promising path to address this challenge is to use global microwave driving on the NV $|\pm1\rangle\leftrightarrow|0\rangle$ transition so that individual control can be performed on the $|+1\rangle\leftrightarrow|-1\rangle$ transition which has a $\sim 50\times$ stronger coupling to the electric field. NV centers also show large spectral diffusion in nanostructures \cite{rodgers2021materials,faraon2012coupling} but several recent works have improved the performance \cite{unknown,chakravarthi2021impact}. Generating large arrays of NV centers with lifetime-limited optical coherence remains an open challenge. 

Beyond NV centers in diamond, the eFPSA architecture applies to other color centers in diamond and other wide-bandgap materials \cite{koehl2011room,sipahigil2016integrated,de2021investigation}, though different properties (e.g., different electric field-spin dipole coupling constants) would require different trade-offs. Group-IV centers in diamond have drawn interest due to their inversion symmetry and optical properties \cite{bradac2019quantum}. However the spin-electric field coupling of these CCs has not been reported. Emitters in silicon carbide have been demonstrated with a large tuning range by electric field, which can be a promising candidate for eFPSA \cite{falk2014electrically}. 


Here we have presented an eFPSA architecture that addresses several challenges in the development of scalable quantum networks. We showed that electric field control is beneficial for the individual quantum addressing of dense emitter arrays, as power consumption and crosstalk are significantly reduced in comparison to the magnetic field case. Furthermore, the wide tunability via the Stark effect allows for multi-channel, parallelized optical entanglement schemes that offer improved scaling with number of qubits. Based on these advantages, we expect that eFPSA architectures will form the basis of future quantum networking implementations.

\section{\label{sec:level1}Acknowledgements\protect}

H.W. acknowledges support from the National Science Foundation Center for Ultracold Atoms (NSF CUA). M.E.T. acknowledges support through the Army Research Laboratory ENIAC Distinguished Postdoctoral Fellowship. L.K. acknowledges support through an appointment to the Intelligence Community Postdoctoral Research Fellowship Program at the Massachusetts Institute of Technology, administered by Oak Ridge Institute for Science and Education through an interagency agreement between the U.S. Department of Energy and the Office of the Director of National Intelligence. D.R.E. acknowledges support from the Bose Research Fellowship, the Army Research Office Multidisciplinary University Research Initiative (ARO MURI) biological transduction program, and the NSF CUA. We thank Ian Christen, Dr. Hyeongrak Choi, Kevin C. Chen and Dr. Lorenzo de Santis for helpful discussions.

\nocite{*}

\bibliography{apssamp}
\appendix

\setcounter{figure}{0}
\renewcommand\thefigure{A\arabic{figure}}

\section{Electric field confinement by dielectric fins}

In the main text, Fig.~2(a) shows the electric field profile of the eFPSA which uses fin structure to confine electric field. For a comparison, we show the field profile without dielectric guiding fins in Fig. A1. The field maximum and the spatial refinement are both reduced by a factor of 1.7, resulting a cross-talk fidelity of $F=0.66$. 

\begin{figure}
\includegraphics[width = 0.47\textwidth]{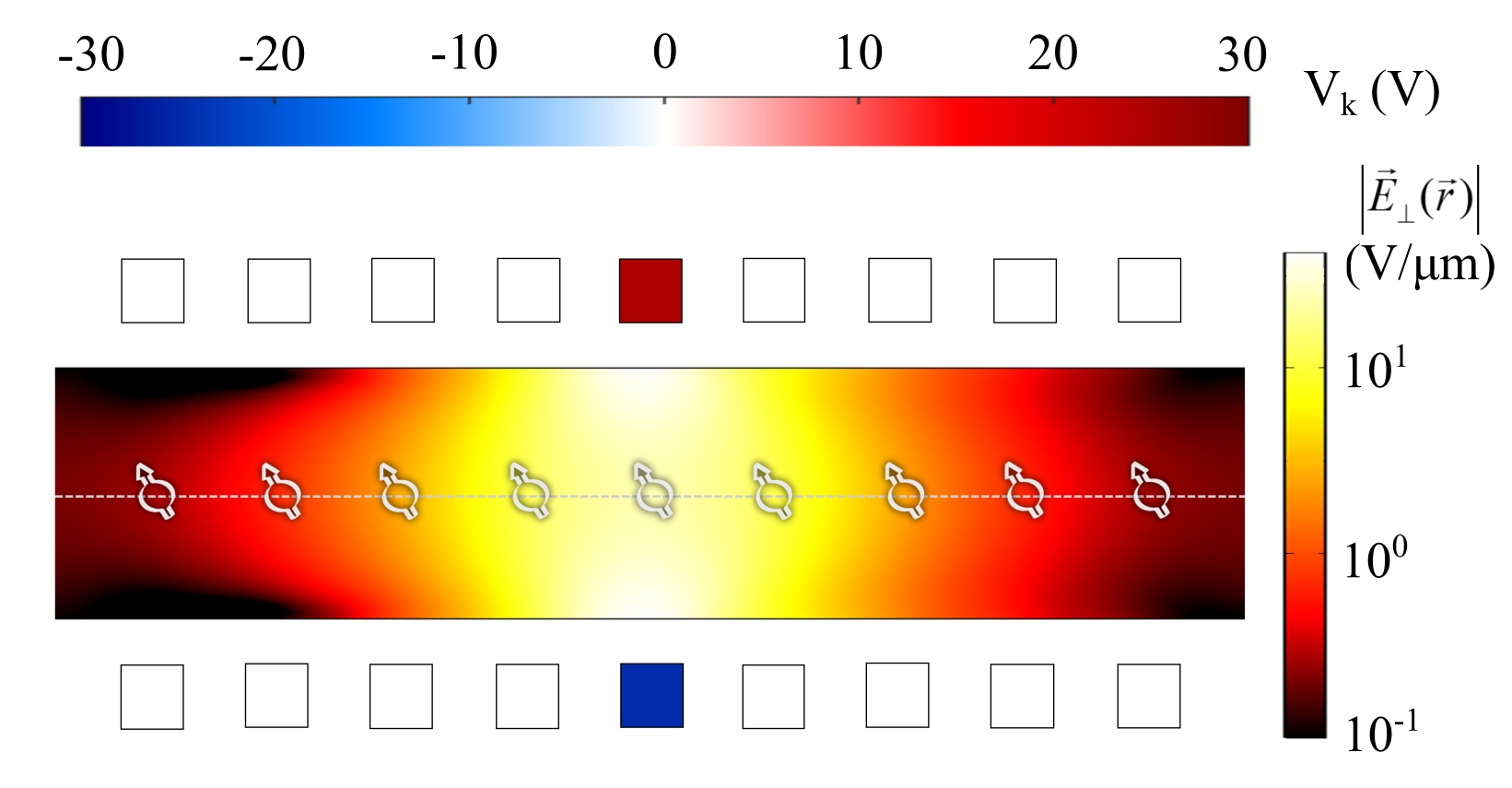}
\caption{\label{fig:A1} Electric field profile without dielectric fins}
\end{figure}

\section{Equivalent circuit of eFPSA}
The eFPSA can be modeled as a capacitor $C$ and a parallel resistance $R$. We neglect the inductance because $\omega L\ll 1/\omega C$ in the eFPSA. We consider a circuit shown in Fig. A2(a), where a voltage source $U$ and the eFPSA are connected by a transmission line with length $l=l_1+l_2$. Here $l_1$ ($l_2$) is the length of transmission line in low temperature (room temperature) part. The low temperature part contains an eFPSA, a series resistance $R_w$ with a parallel wire capacitor $C_w\sim$ fF/$\upmu$m \cite{bernstein2021freely} and a transmission line with length $l_2\ll c/f\sim 0.15$ m. The room temperature (RT) part contains a transmission line with length $l_1\gg l_2$ and a voltage source $U$. 

\begin{figure}[b]
\includegraphics[width = 0.48\textwidth]{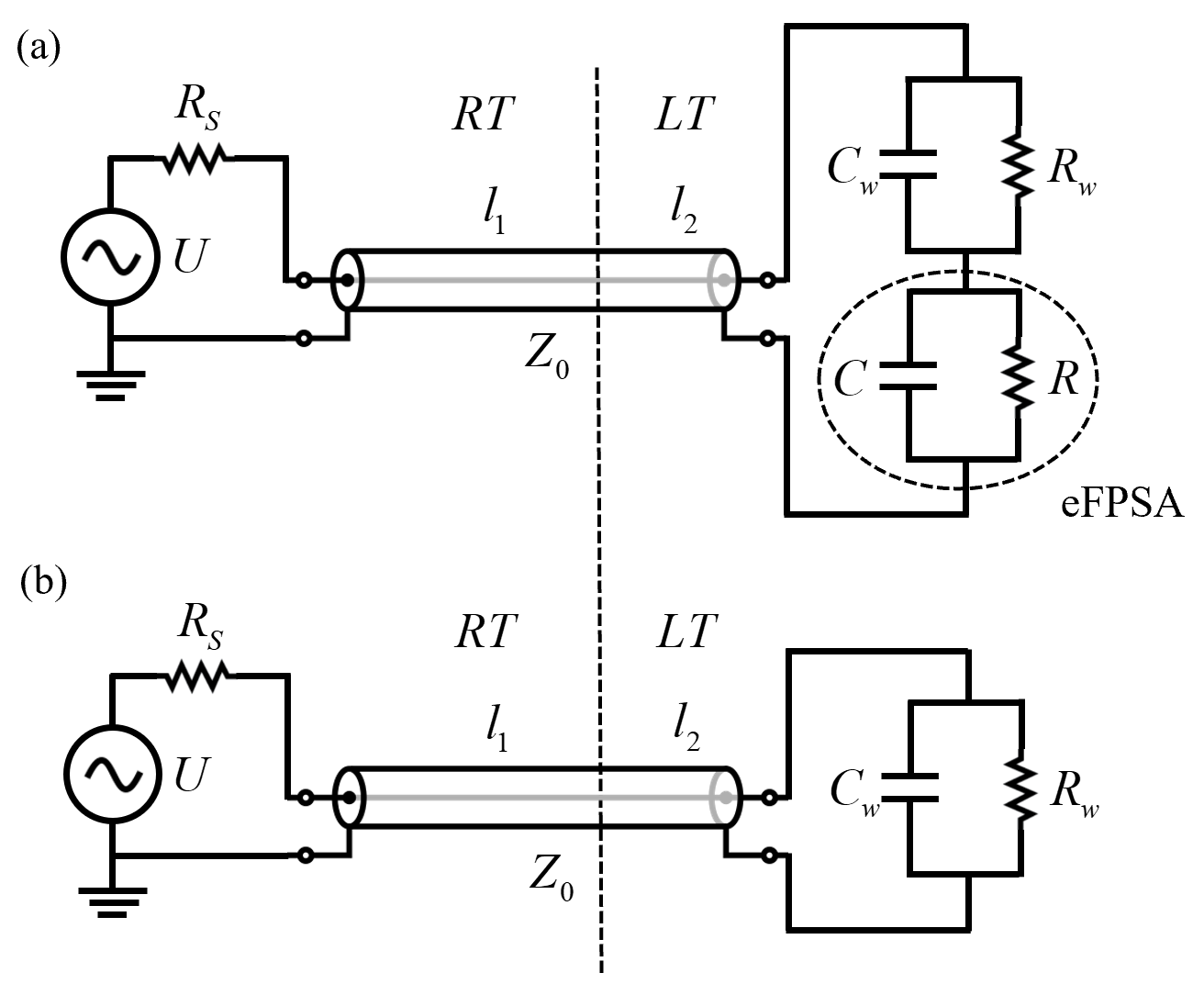}
\caption{\label{fig:A2} (a). Equivalent circuit of eFPSA. Room temperature (RT) part contains a transmission line with length $l_1$ and a voltage source $U$. Low temperature (LT) part contains a transmission line with length $l_2$ and eFPSA. (b). Equivalent circuit for magnetic field-based control. }
\end{figure}

In the following calculation, we omit the wire capacitor $C_w$ because $R_w\ll 1/j\omega C_w$. The impedance of the low-temperature part is:
\begin{equation}
    Z_{LT}=Z_0\frac{Z_C+jZ_0\tan(\beta l_2)}{Z_0+jZ_C\tan(\beta l_2)}\approx Z_C
\end{equation}
where $Z_0=50~\Omega$, $\beta=2\pi/\lambda$ the wavenumber, and
\begin{equation}
    Z_{C}=\frac{R(1-j\omega CR)}{1+\omega^2C^2R^2}+R_w
\end{equation}
Similar to the open circuit, the potential difference across the eFPSA is:
\begin{equation}
    U_{e}=\frac{2Z_C}{Z_C+Z_0}U\sim 2U
\end{equation}
where $U_0$ is the source voltage. The voltage on eFPSA is $2U_0$ and the current is almost zero.

The heat load per $\pi$-pulse in the low temperature is:
\begin{equation}
    J_{E} = (\frac{U_e^2}{R}+|\frac{U_e}{1/j\omega C}|^2R_w)\frac{1}{2\Omega_R}
\end{equation}
with
\begin{equation}
   U_e = \frac{\Omega_R\Lambda}{d_\perp}
\end{equation}
then we have:
\begin{equation}
    J_{E} = \frac{1+\omega^2C^2R_wR}{R}\frac{\Lambda^2\Omega_R}{2d_\perp^2}
\end{equation}
Then we consider a circuit shown in Fig.~A2(b) for magnetic field-based spin control. Here we model the low temperature part as a wire capacitor $C_w$ and a parallel resistance $R_w$, for the same reason we omit the wire capacitor $C_w$ in the calculation. The heat load per $\pi$-pulse in the low temperature part for this circuit is:
\begin{equation}
    J_{B} = \frac{2\pi^2}{\mu_0^2\gamma^2} d^2 R_w\Omega_R
\end{equation}
where $d$ is the distance between wire and NV centers. Here we assume $\Lambda=d$ then we have:
\begin{equation} \label{eq:5}
    \frac{J_E}{J_B}\sim\frac{\mu_0^2\gamma^2}{4\pi^2d_\perp^2}\frac{1+\omega^2C^2R_wR}{R_wR}
\end{equation}

\begin{figure}[h!]
\includegraphics[width = 0.47\textwidth]{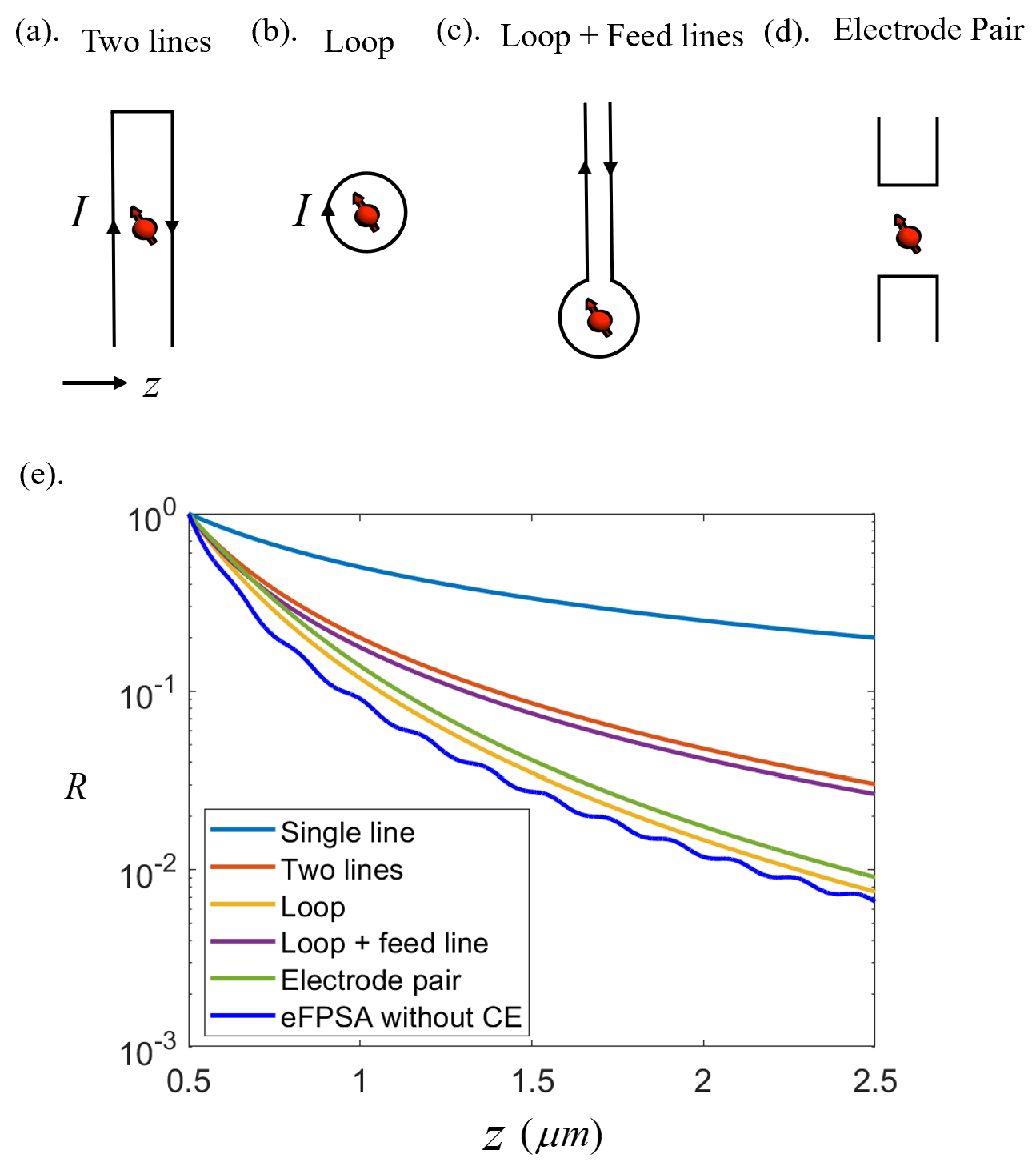}
\caption{\label{fig:A3} (a). Two lines structure, where NVs are set between two lines. (b). Loop structure, where NVs are set in the center of the loops. (c). Loop structure with feed lines. (d). Field profile for all structures normalized by the field at 500 nm.}
\end{figure}

\section{Comparison of field localization between magnetic field-based spin driving and eFPSA}

\begin{figure*}
\includegraphics[width = 0.95\textwidth]{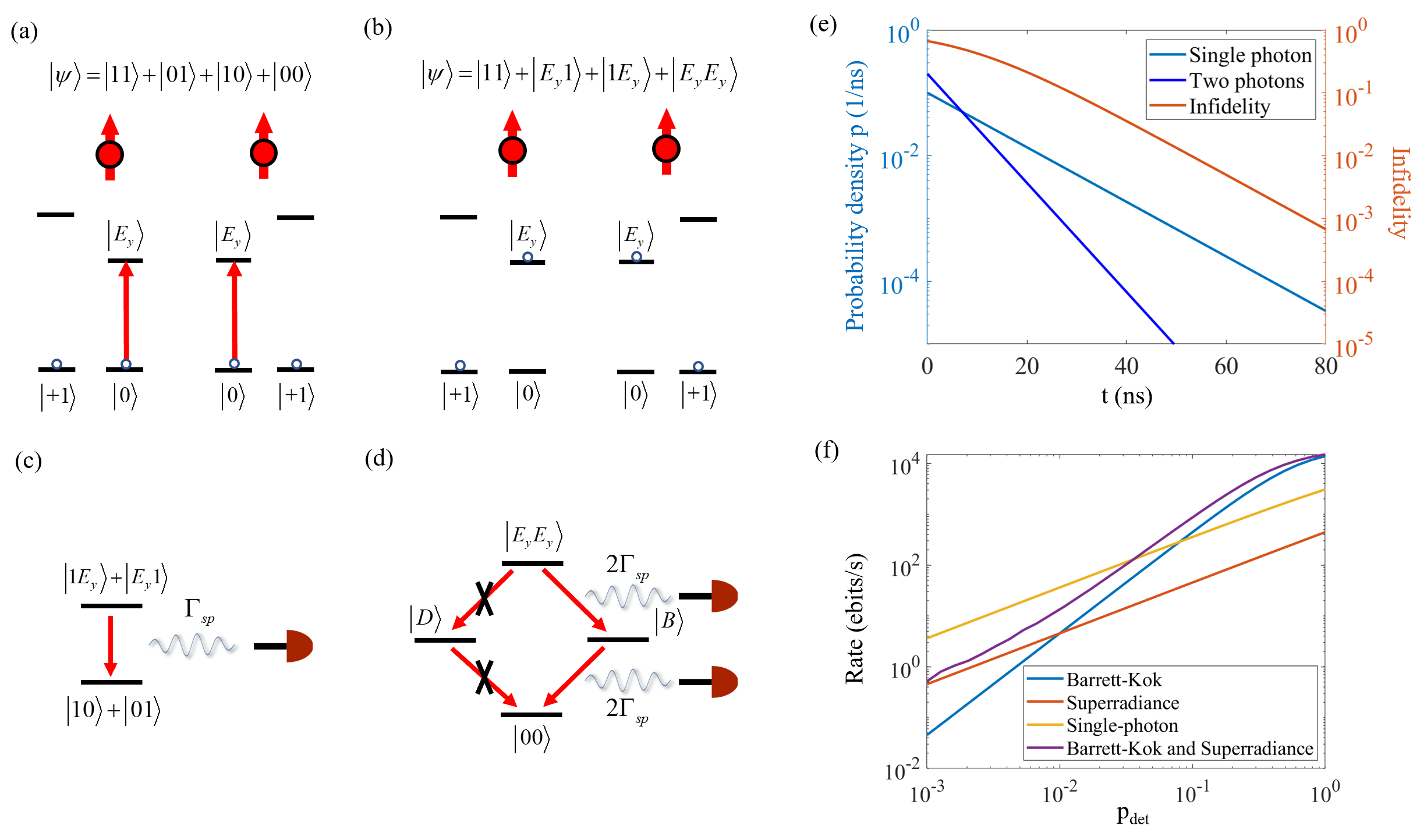}
\caption{\label{fig:A4} Entanglement generation between two NVs. (a). Two NVs are prepared at state $|\Psi\rangle=|11\rangle+|01\rangle+|10\rangle+|00\rangle$. (b). Both NVs are optically excited to $|\Psi\rangle=|11\rangle+|E_y1\rangle+|1E_y\rangle+|E_yE_y\rangle$. (c). Superradiance spontaneous emission with rate $2\Gamma_{sp}$ when NV is at $|E_yE_y\rangle$. (d). Single photon spontaneous emission with rate $\Gamma_{sp}$ when NV is in $|1E_y\rangle+|E_y1\rangle$. (e). Probability for emitting a photon at time $t$ for cases (c) and (d) and the infidelity to distinguish both cases. (f). Entanglement rate for Barrett-Kok scheme, single-photon scheme, superradiance and a combination of superradiance and Barrett-Kok scheme. }
\end{figure*}

We compare the localization of electric field and magnetic field based on example geometries to derive the scaling laws. In the magnetic field case we consider the single line, two lines, loop, loop and feed lines shown in Fig. A3(a-c). In the electric field case, we consider electrode pair and eFPSA shown in Fig. A3(d) and Fig. 1(a). Here we assume the distance between the NV and nearest structure is $a= 250~$nm for all geometries, and normalize the field at $z=500$ nm for comparison.

We show the calculated field profiles in Fig. A3(c), using Biot-Sarvart law for magnetic field and COMSOL simulation for electric field geometries. For a single wire, magnetic field falls slowly with $B\propto 1/r$. Structures with two opposing currents will locally cancel the magnetic field. The magnetic field from these structures (two-lines and loop) falls off as $B\propto 1/r^2$ and $B\propto 1/r^3$. However, the loop may need to be connected by feed lines for transferring current shown in Fig. A3(c). Then the scaling then will be limited by $B\propto 1/r^2$. In the electric field case, we consider an electrode pair, the electric field will scale as $E \propto 1/r^3$. For eFPSA, despite the many electrodes connected to the ground to help localize the electric field, the electric field still scales as $E \propto 1/r^3$.

\section{Entanglement generation with superradiance in a photonic crystal waveguide}

Here we consider the entanglement scheme for an NV pair in a photonic crystal waveguide. We prepare two NVs in $|\Psi\rangle=|11\rangle+|01\rangle+|10\rangle+|00\rangle$ and excite $|0\rangle$ to $|E_y\rangle$, as shown in Fig. A4(a) and Fig. A4(b). We then treat the spontaneous emission of each component of $|\Psi\rangle=|11\rangle+|E_y1\rangle+|1E_y\rangle+|E_yE_y\rangle$ individually. The $|11\rangle$ state will remain unchanged under optical emission timescales. The $|E_y1\rangle$ or $|1E_y\rangle$ states will emit a single photon with spontaneous emission rate $\Gamma_{sp}=100$ MHz \cite{chu2014coherent} (Fig. A4(c)), where we assume a Purcell factor of 10. Due to the superradiance effect in the photonic crystal waveguide, the $|E_yE_y\rangle$ state will first radiatively decay with rate $2\Gamma_{sp}$ to the bright Dicke state $|B\rangle=|E_y0\rangle+e^{ikL}|0E_y\rangle$, where $L$ is the distance between two NVs and $k$ is the photon wavevector \cite{kim2018super}. Then another photon is emitted when the Dicke state decays to the ground state at rate $2\Gamma_{sp}$. The decay rate of $2\Gamma_{sp}$ is the critical difference between independent emission and superradiant emission in the photonic crystal waveguide.


In the quantum entanglement generation scheme, the main infidelity is misheralding. Upon single photon detection, the desired state is $|\Psi_0\rangle=|01\rangle+|10\rangle$ following from the $|E_y1\rangle + |1E_y\rangle$ state. However, a single photon can also be detected after the spontaneous emission by $|E_yE_y\rangle$ and subsequent loss, in which case the state is $|00\rangle$ and misheralding has occurred. In the single photon scheme \cite{humphreys2018deterministic}, the probability of $|E_yE_y\rangle$ is reduced by preparing a superposition state $|\alpha\rangle=\sqrt{\alpha}|0\rangle+\sqrt{1-\alpha}|1\rangle$. By choosing $\alpha$ factor, the fidelity $F=1-\alpha$ will be traded-off with entanglement rate $r=2\alpha p_{det}$. In Barrett-Kok scheme \cite{barrett2005efficient, bernien2013heralded}, this error is eliminated by flipping the spin and repeating the optical heralding process, where a second photon detection is not possible from the $|11\rangle$ state. This two-step scheme leads to an entanglement rate $r=p_{det}^2/2$.

The change of the lifetime in a slow-light waveguide offers a potential path to decrease the infidelity due to misheralding. Since the emission from the desired state has a longer lifetime, we can differentiate superradiance and standard cases. Specifically, the detection of a single photon at $t$ results in a fidelity:

\begin{equation}
    F = \frac{\exp(\Gamma_{sp}t)}{2+\exp(\Gamma_{sp}t)}
\end{equation}

However, the probability density for detecting a photon at $t\rightarrow t+dt$ is:
\begin{equation}
    p(t) = \Gamma_{sp}\exp(-\Gamma_{sp}t)
\end{equation}
Finally the rate-fidelity trade-off for this process can be written as:
\begin{equation}
    p = \frac{1-F}{2F}
\end{equation}

As shown in Fig. A4(e), the photon detected after $t_0=53$ ns heralds the entanglement generation with a fidelity $F>0.99$. The probability to get a photon after $t_0=53$ ns is $p= 5\times 10^{-3}$. We compare the rate using several schemes with $F=0.99$ in Fig. A4(f). Used alone, the time-domain filtering doesn't give an advantage over other scehmes. However, timing information can be recorded in conjunction with other schemes. For example, a photon detection at $t>t_0$ already heralds high-fidelity entanglement in a Barret-Kok scheme, rendering the second heralding step unnecessary for eliminating heralding error. The combination of Barrett-Kok scheme and superradiance gives the best rate when photon detection efficiency $p_{det}>3\times 10^{-2}$.

\end{document}